\documentclass[sigconf,unicode,bookmarks=false, natbib=false]{acmart}
\PassOptionsToPackage{bookmarks=false}{hyperref}
\usepackage{xcolor}
\usepackage[inline]{enumitem}
\usepackage{verbatim}
\usepackage{booktabs} % For formal tables
\usepackage{tabularx}
\usepackage{xcolor,colortbl}
\usepackage{comment}
\usepackage{enumitem}
\usepackage{soul}
\usepackage{outlines}

\usepackage[most]{tcolorbox}
\usepackage{fontawesome}
\definecolor{boxcolor}{RGB}{238, 223, 204} %
\DeclareRobustCommand{\mybox}[2][gray!5]{%
\begin{tcolorbox}[   %% Adjust the following parameters at will.
        breakable,
        left=0pt,
        right=0pt,
        top=0pt,
        bottom=0pt,
        colback=#1,
        colframe=black,
        width=\dimexpr\columnwidth\relax, 
        enlarge left by=0mm,
        boxsep=5pt,
        outer arc=4pt,
        boxrule=.5mm
        ]
        #2
\end{tcolorbox}
}

\usepackage{tikz}
\usepackage{pgfplots}
\pgfplotsset{width=10cm,compat=1.9}

\newlength{\myboxheight}
\usepackage[style=ACM-Reference-Format]{biblatex}
\acmYear{2022}
\acmConference[CHASE'22]{15th International Conference on Cooperative and Human Aspects of Software Engineering}{May 23, 2022}{Pittsburgh, PA, USA}
\acmDOI{}
\acmISBN{}

\setcopyright{none}

\begin{document}

\title[Communication as a Resource]{``Communication Is a Scarce Resource!''}
\subtitle{A Summary of CHASE'22 Conference Discussions}


\author{Christoph Matthies}
\affiliation{
      \institution{Hasso Plattner Institute\\University of Potsdam}
  \streetaddress{August-Bebel-Str. 88, 14482 Potsdam}
  \city{Potsdam}
  \country{Germany}}
\email{christoph.matthies@hpi.de}

\author{Mary Sánchez-Gordón}
\affiliation{
  \institution{Østfold University College}
  \streetaddress{BRA Veien 4, 1757}
  \city{Halden}
  \country{Norway}}
\email{mary.sanchez-gordon@hiof.no}

\author{Jens Bæk Jørgensen}
\affiliation{
  \institution{Mjølner Informatics A/S}
  \streetaddress{Finlandsgade 10}
  \city{Aarhus N}
  \country{Denmark}}
\email{jbj@mjolner.dk}

\author{Lutz Prechelt}
\affiliation{
  \institution{Freie Universität Berlin}
  \streetaddress{Takustr. 9}
  \city{Berlin}
  \country{Germany}}
\email{prechelt@inf.fu-berlin.de}

% TODO
\renewcommand{\shortauthors}{Matthies, Sánchez-Gordón, Jørgensen, Prechelt}

\begin{abstract}
\noindent\textbf{Background:}
Software Engineering regularly views communication between project participants as a tool for solving various problems in software development.

\noindent\textbf{Objective:} 
Formulate research questions in areas related to CHASE.

\noindent\textbf{Method:}
A day-long discussion of five participants at the in-person day of 
the \textit{15th International Conference on Cooperative and Human Aspects of Software Engineering} (CHASE 2022) on May 23rd 2022.

\noindent\textbf{Results:}
It is not rare in industrial SE projects that communication is not just a tool or technique to be applied but also represents a resource, which, when lacking, threatens project success. This situation might arise when a person required to make decisions (especially regarding requirements, budgets, or priorities) is often unavailable.
It may be helpful to frame communication as a scarce resource to understand the key difficulty of such situations.

\noindent\textbf{Conclusion:}
We call for studies that focus on the allocation and management of scarce communication resources of stakeholders as a lens to analyze software engineering projects.

\end{abstract}

%   <concept>
%       <concept_id>10011007.10011074.10011081.10011082.10011083</concept_id>
%       <concept_desc>Software and its engineering~Agile software development</concept_desc>
%       <concept_significance>500</concept_significance>
%       </concept>
%  </ccs2012>

% \ccsdesc[500]{Software and its engineering~Agile software development}


%   \caption{Caption.}
%   \Description{Description.}
\maketitle

\section{Introduction}

The 15th CHASE conference took place in May 2022 in a hybrid model: as an online event on May 18-19th and an in-person event on May 23rd, colocated with the ICSE 2022 conference in Pittsburgh~\cite{chaseweb}.
%\footnote{https://conf.researchr.org/home/chase-2022}.
The in-person CHASE 2022 conference was an interactive workshop, with the attendees forming two ad-hoc work groups.
These groups tasked themselves with formulating research questions regarding cooperative and human factors in software engineering.

\subsection{Work Group Participants}
This report summarizes the discussion outcomes of the CHASE'22 work group, including Lutz Prechelt, Christoph Matthies, Jens Bæk Jørgensen, Mary Sánchez-Gordón, and Gennadiy Civil.

\begin{figure}[h]
    \centering
    \includegraphics[width=0.95\columnwidth]{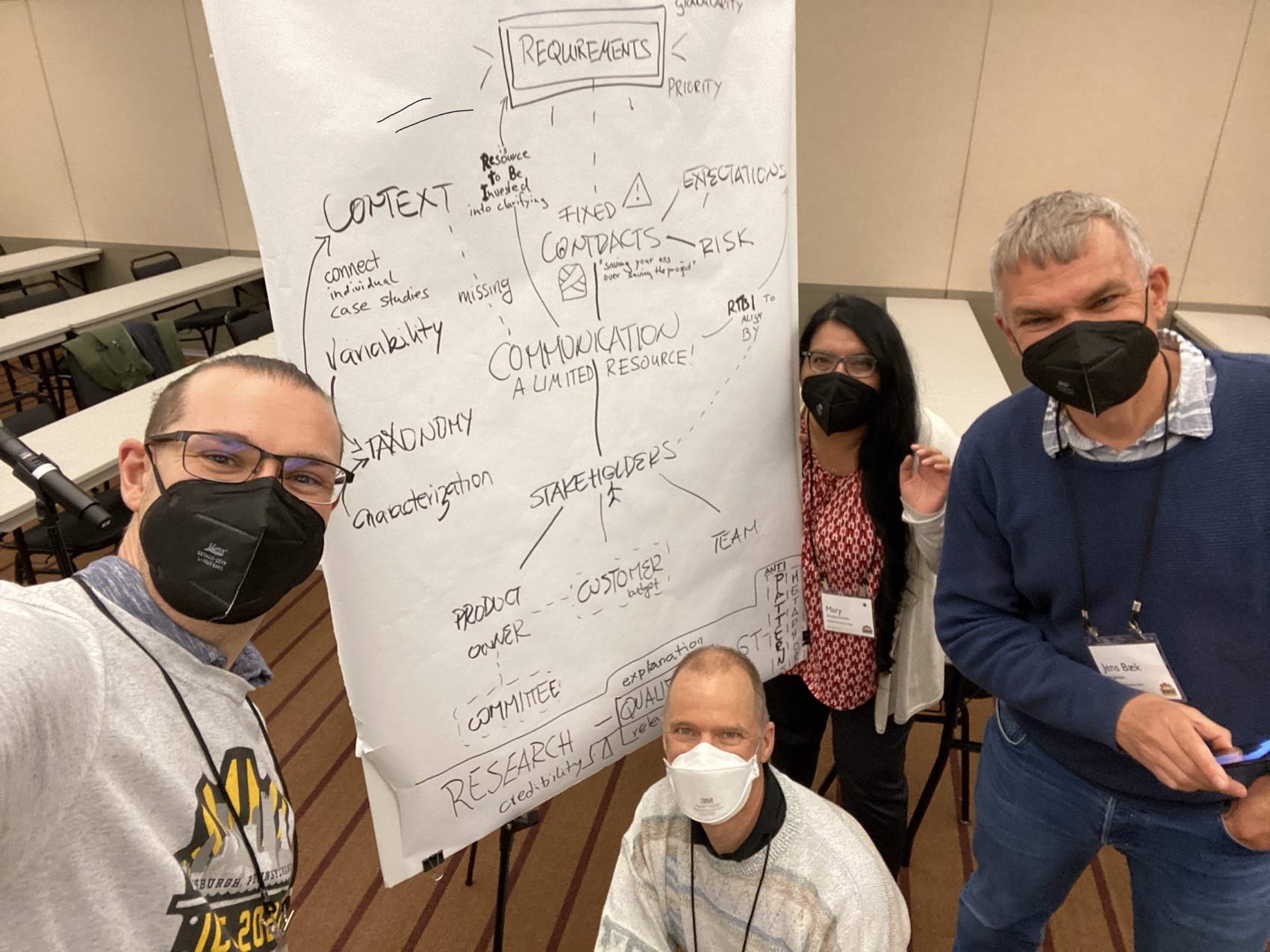}
    \caption{CHASE'22 working group participants and a mind map of the discussed topics}
    \label{fig:selfie}
\end{figure}

Participants summarized their previous experiences in academic and software industry settings, revealing that Jens was the most active software professional in the group as a senior project manager at a Danish software consultancy~\cite{mjolnerweb}.
%\footnote{https://mjolner.dk/en/}.
Through his accounts, the groups gained insights into the varied current challenges related to the human aspects within the company's agile software processes.
The other group participants contributed related experiences and focused on understanding the issues' root causes through follow-up questions.
We structured the group's findings in the form of a visual mind map, see Figure~\ref{fig:selfie}.

\subsection{Work Group Topics}
The work group focused on the challenges of using agile methods at scale, particularly in consulting activities and managing multiple agile teams.
These included:
\begin{itemize}
    \item Contract negotiation in the context of agile development
    \item Project stakeholder alignment
    \item The Product Owner (PO) role in complex industry projects
\end{itemize}

Our discussion then focused on the underlying process issues and root causes of the presented challenges.
We considered future research directions and methods to tackle the identified problems and came to insights in two main topic areas:
\begin{itemize}
    \item \textbf{Software process issues}. Characterization of the communication capacity of project stakeholders as a scarce resource within organizations.
    \item \textbf{Qualitative research methods}. Call to move from descriptive research to explanatory theories of behavior patterns.
\end{itemize}

\subsection{Contributions}

\noindent We summarized the group's discussion under a single headline:
\noindent \emph{``Communication is sometimes a scarce resource in software development projects!''}

\smallskip
We envision this guiding principle to be helpful as a framework to think about and investigate software project setups.
The scarce resource of communication is required in all development contexts, especially in agile methods with a focus on individuals and interactions and customer collaboration.
In fact, in the Agile Manifesto, only four of the twelve principles are formulated as imperatives and the only one that talks about the amount of required communication is among them: 
\emph{``Business people and developers must work together daily throughout the project.''}~\cite{beck2001manifesto}

However in practice, every process participant has a limited supply of communication resources, which require careful, considerate management and prioritization, just as other project resources and budgets.

Framing communication as a resource to be invested by stakeholders, rather than an outcome of assigned responsibilities, enables a new perspective on development processes that takes investments and returns into explicit account.

\mybox{\faArrowCircleRight~\textbf{Takeaway:}
    We propose the guiding principle of \textbf{``Communication is a scarce resource''} in software development as a lens to investigate development setups.
    As communication resources of stakeholders are limited, they require careful management.
}

\section{Process and Communication Challenges}
\label{sec:communication}

This section summarizes the discussed communication challenges in agile software processes.

\subsection{Product Owners in Large-Scale Development}
After reflecting on the root causes, the role of the Product Owner in complex, large-scale agile software development projects emerged as a hard-to-fill position. Although the Scrum Guide states: ``the Product Owner is one person, not a committee'', it also specifies that ``the Product Owner may do the above work or may delegate the responsibility to others''~\cite{schwaber2020}. In practice, the PO is involved in multiple vital parts of the software development process and needs to communicate with various stakeholders. In sufficiently large and complex industry settings, this large workload needs to be handled by multiple people, e.g., a \emph{committee}. This raises the question: \emph{how can this tension between the undesirable ``PO committee'' and the required delegation of responsibility be navigated?}

\subsection{Fixed-Price Contracts in Agile Projects}
Fixed-price contracts are agreements to complete projects for a flat rate, i.e., the promise of delivering a product for a given cost, time, and scope. In this case, we highlighted the perception that projects using such contracts focus on shifting responsibilities to others rather than delivering business value. Therefore, fixed-price contracts were seen as an ``anti-pattern'' to be avoided as much as possible in agile contexts. We also recognized that alternative approaches, such as buying work instead of features, were still challenging to communicate to industry clients.

\subsection{Communication as a Project Resource}
We used a mind map (see Figure~\ref{fig:selfie}) to structure our collected insights around a central concept: \emph{communication as a ``project resource''}. 
We identified that communication is particularly relevant in aligning stakeholder expectations and clarifying project requirements.
We labeled these communication relationships in our visual representation as a \textbf{``Resource To Be Invested''} (RTBI), i.e., communication is an RTBI to refine project requirements.

This framing highlights both the inherent trade-off in resource allocation as well as the need for resource management. Assigning additional communication tasks to a role leaves that role with fewer resources for other communication duties. Suppose more communication resources are to be invested in a particular project area by the project stakeholders. In that case, this must lead to a decrease somewhere else. Moreover, although communication resources might be expanded, i.e., by involving more stakeholders, this will result in increased synchronization overhead. Therefore, it is not sufficient to call for additional communication to solve inherent process issues!

\mybox{\faArrowCircleRight~\textbf{Takeaway:}
    Given that communication resources in software projects are scarce, communication can be framed as a key \textbf{``Resource To Be Invested''} (RTBI) by project stakeholders.
    Investing communication resources involves trade-offs between different investments and their (anticipated) returns.
}

\section{Thoughts on Suitable Research Approaches}
\label{sec:qual_research}

Our work group discussed research methods to address the identified issues regarding three aspects:

\begin{itemize}
    \item \textbf{Method Types.} Qualitative research methods, in particular the application of Grounded Theory, are well-suited to better understand the specifics of the raised software process issues.
    \item \textbf{Quality Criteria.} Credibility and relevance as the main aspects of assessing the impact of research studies and as selection criteria for analyses.
    \item \textbf{Theories.} Moving from basic descriptions of human behavior observed in studies to identifying explanatory theories derived from multiple sources benefits the research field.
\end{itemize}

\subsection{Towards Named SE (Anti-)Patterns}

A key group discussion outcome is a call to the software engineering research community to produce well-defined (anti-)patterns within software projects and team communications.
Ideally, these anti-patterns are named and include metaphors, as is the case with software design patterns, such as \emph{Bridge}, \textit{Factory Method} or \emph{Observer}.
These names help facilitate communication by researchers as well as practitioners.

However, these metaphors must be chosen carefully.
As an example of the ill effects of selecting an analogy, we discussed the term ``technical debt''.
While the concept might not have been designed to be quantified, the notion of ``monetary debt'' in terms of currency owed made assigning measurements a seemingly logical next step, which probably has not contributed to the usefulness of the metaphor.

\subsection{Towards Connecting Case Studies}

The group also highlighted the need for connecting different case studies with the goal of developing explanatory theories in software engineering.

Vigorous qualitative research studies to build theories are complex and time-consuming projects.
These might not necessarily be possible in every desired circumstance.
In these cases, we call on a focus on accurate and detailed descriptions of the study context so that results can be used and interpreted in following studies.
A taxonomy of SE study context properties would allow characterizing the context so that researchers can combine descriptions of multiple case studies into more extensive theories regardless of their variability.
We call on practitioners and academics to collaborate on case studies or experience reports that accurately present the case study context and have both high relevance and good-enough credibility. 

Our work group discussions highlighted that practitioners have real-world experiences they wish to share with the broader community.
However, they may not have sufficiently strong knowledge of sound research methods and scientific writing to make a contribution with high credibility; their descriptions may get too anecdotal to provide scientific value.
Moreover, practitioners often do not have an overview of relevant and current related work to put their experiences in the proper perspective, i.e., identifying what is new or different in their current context from what has already been described and published.

\mybox{\faArrowCircleRight~\textbf{Takeaway:} 
    Qualitative studies that lead to explanatory theories in software engineering are still lacking.
    \textbf{Individual use case studies can contribute to a larger body of knowledge by providing detailed, structured context descriptions.}
    An agreed-upon taxonomy of context descriptions in SE would allow connecting case study results to analyze the evidence of multiple studies in the presence of variability.
}

% Balance columns with references
\balance

% \bibliographystyle{ACM-Reference-Format}
% \bibliography{bib}
\printbibliography

\end{document}